\documentclass[nameyear,seceqn,dvips]{arxstspdf}
\usepackage{flushend}
\usepackage{stfloats}

\volume{26}
\issue{2}
\pubyear{2011}
\firstpage{187}
\lastpage{202}
\doi{10.1214/10-STS338}

\makeatletter
\newtheorem{theorem}{Theorem}
\newproclaim{remark}{Remark}
\newproclaim{example}{Example}
\newcommand{\ra}{\rightarrow}

\newcommand{\la}{\lambda}
\newcommand{\bet}{\beta}
\renewcommand{\th}{\theta}
\makeatother

\begin{document}
\begin{frontmatter}

\title{Objective Priors: An Introduction for Frequentists\thanksref{T1}}
\relateddoi{T1}{Discussed in \doi{10.1214/11-STS338A} and
\doi{10.1214/11-STS338B}; rejoinder at \doi{10.1214/11-STS338REJ}.}

\runtitle{Objective Priors}

\begin{aug}
\author{\fnms{Malay} \snm{Ghosh}\corref{}\ead[label=e1]{ghoshm@stat.ufl.edu}}

\runauthor{M. Ghosh}

\affiliation{University of Florida}

\address{Malay Ghosh is Distinguished Professor, University of Florida, 223
Griffin--Floyd Hall, Gainesville, Florida 32611-8545, USA (e-mail:
ghoshm@stat.ufl.edu).}

\end{aug}

%
\begin{abstract}
Bayesian methods are increasingly applied in these days in the theory
and practice of statistics. Any Bayesian inference depends on a
likelihood and a prior. Ideally one would like to elicit a prior from
related sources of information or past data. However, in its absence,
Bayesian methods need to rely on some ``objective'' or ``default''
priors, and the resulting posterior inference can still be quite
valuable.

Not surprisingly, over the years, the catalog of objective priors also
has become prohibitively large, and one has to set some specific
criteria for the selection of such priors. Our aim is to review some of
these criteria, compare their performance, and illustrate them with
some simple examples. While for very large sample sizes, it does not
possibly matter what objective prior one uses, the selection of such
a prior does influence inference for small or moderate samples. For
regular models where asymptotic normality holds, Jeffreys' general rule
prior, the positive square root of the determinant of the Fisher
information matrix, enjoys many optimality properties in the absence of
nuisance parameters. In the presence of nuisance parameters, however,
there are many other priors which emerge as optimal depending on the
criterion selected. One new feature in this article is that a prior
different from Jeffreys' is shown to be optimal under the chi-square
divergence criterion even in the absence of nuisance parameters. The
latter is also invariant under one-to-one reparameterization.
\end{abstract}

%
\begin{keyword}
\kwd{Asymptotic expansion}
\kwd{divergence criterion}
\kwd{first-order probability matching}
\kwd{Jeffreys' prior}
\kwd{left Haar priors}
\kwd{location family}
\kwd{location--scale family}
\kwd{multiparameter}
\kwd{orthogonality}
\kwd{reference priors}
\kwd{right Haar priors}
\kwd{scale family}
\kwd{second-order probability matching}
\kwd{shrinkage argument}.
\end{keyword}

\end{frontmatter}

\section{Introduction}\label{s1}

Bayesian methods are increasingly used in recent years in the theory
and practice of statistics. Their implementation requires specification
of both a likelihood and a prior. With enough historical data, it is
possible to elicit a prior distribution fairly accurately. However,
even in its absence, Bayesian methods, if judiciously used, can produce
meaningful inferences based on the so-called ``objective'' or
``default'' priors.

The main focus of this article is to introduce certain objective priors
which could be potentially useful even for frequentist inference. One
such example where frequentists are yet to reach a consensus about an
``optimal'' approach is the construction of confidence intervals for
the ratio of two normal means, the celebrated Fieller--Creasy problem.
It is shown in Section \ref{s4} of this paper how an ``objective''
prior produces a credible interval in this case which meets the target
coverage probability of a frequentist confidence interval even for
small or moderate sample sizes. Another situation, which has often
become a real challenge for frequentists, is to find a suitable method
for elimination of nuisance parameters when the dimension of the
parameter grows in direct proportion to the sample size. This is what
is usually referred to as the Neyman--Scott phenomenon. We will
illustrate in Section \ref{s3} with an example of how an objective
prior can sometimes overcome this problem.

Before getting into the main theme of this paper, we recount briefly
the early history of objective priors. One of the earliest uses is
usually attributed to Bayes (\citeyear{b1763}) and Laplace
(\citeyear{l1812}) who recommended using a uniform prior for the
binomial proportion $p$ in the absence of any other information. While
intuitively quite appealing, this prior has often been criticized due
to its lack of invariance under one-to-one reparameterization. For
example, a uniform prior for $p$ in the binomial case does not result
in a uniform prior for $p^{2}$. A~more compelling example is that a
uniform prior for $\sigma$, the population standard deviation, does not
result in a uniform prior for $\sigma^{2}$, and the converse is also
true. In a situation like this, it is not at all clear whether there
can be any preference to assign a uniform prior to either $\sigma$ or
$\sigma^{2}$.

In contrast, Jeffreys' (\citeyear{j1961}) general rule prior, na\-mely,
the positive square root of the determinant of the Fisher information
matrix, is invariant under one-to-one reparameterization of parameters.
We\break will motivate this prior from several asymptotic considerations. In
particular, for regular models where asymptotic normality holds,
Jeffreys' prior enjoys many optimality properties in the absence of
nuisance parameters. In the presence of nuisance parameters, this prior
suffers from many problems---marginalization paradox, the Neyman--Scott
problem, just to name a few. Indeed, for the location--scale models,
Jeffreys himself recommended alternate priors.

There are several criteria for the construction of objective priors.
The present article primarily reviews two of these criteria in some
detail, namely, ``divergence priors'' and ``probability matching\vadjust{\eject}
priors,'' and finds optimal priors under these criteria. The class of
divergence priors includes ``reference priors'' introduced by Bernardo
(\citeyear{b1979}). The ``probablity matching priors'' were introduced
by Welch and Peers (\citeyear{wp1963}). There are many generalizations
of the same in the past two decades. The development of both these
priors rely on asymptotic considerations. Somewhat more briefly, I~have
discussed also a few other priors including the ``right'' and ``left''
Haar priors.

The paper does not claim the extensive thorough and comprehensive
review of Kass and Wasserman (\citeyear{kw1996}), nor does it aspire to
the somewhat narrowly focused, but a very comprehensive review of
probability matching priors as given in Ghosh and Mukerjee
(\citeyear{gm1998}), Datta and Mukerjee (\citeyear{dm2004}) and Datta
and Sweeting (\citeyear{ds2005}). A very comprehensive review of
reference priors is now available in Bernardo (\citeyear{b2005}), and a
unified approach is given in the recent article of Berger, Bernardo and
Sun (\citeyear{bbs2009}).

While primarily a review, the present article has been able to unify as
well as generalize some of the previously considered criteria, for
example, viewing the reference priors as members of a bigger class of
divergence priors. Interestingly, with some of these criteria as
presented here, it is possible to construct some alternatives to
Jeffreys' prior even in the absence of nuisance parameters.

The outline of the remaining sections is as follows. In Section
\ref{s2} we introduce two basic tools to be used repeatedly in the
subsequent sections. One such tool involving asymptotic expansion of
the posterior density is due to Johnson (\citeyear{j1970}), and Ghosh,
Sinha and Joshi (\citeyear{gsj1982}), and is discussed quite
extensively in Ghosh, Delampady and Samanta (\citeyear{gds2006}) and
Datta and Mukerjee (\citeyear{dm2004}). The second tool involves a
shrinkage argument suggested by Dawid and used extensively by J.~K.
Ghosh and his co-authors. It is shown in Section~\ref{s3} that this
shrinkage argument can also be used in deriving priors with the
criterion of maximizing the distance between the prior and the
posterior. The distance measure used includes, but is not limited to,
the Kullback--Leibler (K--L) distance considered in Bernardo
(\citeyear{b1979}) for constructing two-group ``reference priors.''
Also, in this section we have considered a new prior different from
Jeffreys even in the one-parameter case which is also invariant under
one-to-one reparameterization. Section \ref{s4} addresses construction
of priors under probability matching criteria. Certain other priors are
introduced in Section \ref{s5}, and it is pointed out that some of
these priors can often provide\vadjust{\eject} exact and not just asymptotic matching.
Some final remarks are made in Section~\ref{s6}.

Throughout this paper the results are presented more or less in a
heuristic fashion, that is, without paying much attention to the
regularity conditions needed to justify these results. More emphasis is
placed on the application of these results in the construction of
objective priors.

\section{Two Basic Tools}\label{s2}

An asymptotic expansion of the posterior density began with Johnson
(\citeyear{j1970}), followed up later by Ghosh, Sinha and Joshi
(\citeyear{gsj1982}), and many others. The result goes beyond that of
the theorem of Bernstein and Von Mises which provides asymptotic
normality of the posterior density. Typically, such an expansion is
centered around the MLE (and occasionally the posterior mode), and
requires only derivatives of the log-likelihood with respect to the
parameters, and evaluated at their MLE's. These expansions are
available even for heavy-tailed densities such as Cauchy because
finiteness of moments of the distribution is not needed. The result
goes a long way in finding asymptotic expansion for the posterior
moments of parameters of interest as well as in finding asymptotic
posterior predictive distributions.

The asymptotic expansion of the posterior resembles that of an
Edgeworth expansion, but, unlike the latter, this approach does not
need use of cumulants of the distribution. Finding cumulants, though
conceptually easy, can become quite formidable, especially in the
presence of multiple parameters, demanding evaluation of mixed
cumulants.

We have used this expansion as a first step in the derivation of
objective priors under different criteria. Together with the shrinkage
argument as mentioned earlier in the \hyperref[s1]{Introduction}, and to be discussed
later in this section, one can easily unify and extend many of the
known results on prior selection. In particular, we will see later in
this section how some of the reference priors of Bernardo
(\citeyear{b1979}) can be found via application of these two tools. The
approach also leads to a somewhat surprising result involving
asymptotic expansion of the distribution function of the MLE in a
fairly general setup, and is not restricted to any particular family of
distributions, for example, the exponential family, or the
location--scale family. A detailed exposition is available in Datta and
Mukerjee (\citeyear{dm2004}, pages 5--8).

For simplicity of exposition, we consider primarily the one-parameter
case. Results needed for\vadjust{\eject} the multiparameter case will occasionally be
mentioned, and, in most cases, these are straightforward, albeit often
cumbersome, extensions of one-parameter results. Moreover, as stated in
the \hyperref[s1]{Introduction}, the results will be given without full rigor, that
is, without any specific mention of the needed regularity conditions.

We begin with $X_{1}, \dots, X_{n}|\theta$ i.i.d. with common p.d.f.
$f(X|\theta)$. Let $\hat{\theta}_{n}$ denote the MLE of $\theta$. The
likelihood function is denoted by $L_{n} (\theta) =
\prod^{n}_{1}f(X_{i}|\theta)$ and let $\ell_{n}(\theta)=\log
L_{n}(\theta)$. Let $a_{i} =
n^{-1}[d^{i}\ell_{n}(\theta)/\break d\theta^{i}]_{\theta= \hat{\theta}_{n}}$,
$i=1,2,\dots,$ and let $\hat{I}_{n} = -a_{2}$, the observed per unit
Fisher information number. Consider a twice differentiable prior $\pi$.
Let $T_{n}=\sqrt{n}(\theta- \hat{\theta}_{n}) \hat{I}_{n}^{1/2}$, and
let $\pi^{*}_{n} (t)$ denote the posterior p.d.f. of $T_{n}$ given\break $X_{1},
\ldots, X_{n}$. Then, under certain regularity conditions, we have the
following result.

\begin{theorem}\label{teo1}
\hspace*{-1pt}$\pi^{*}_{n} (t) = \phi(t) [1+ n^{-1/2} \gamma_{1} (t; X_{1},
\ldots,\break
X_{n}) + n^{-1} \gamma_{2} (t; X_{1}, \dots, X_{n})] + O_{p}
(n^{-3/2})$, where $\phi(t)$ is the standard normal p.d.f., $
\gamma_{1}(t; X_{1}, \dots, X_{n})=\break
a_{3}t^3/(6\hat{I}_{n}^{3/2})+(t/\hat{I}_{n}^{1/2})\pi^{\prime
}(\hat{\theta}_{n})/\pi(\hat{\theta}_{n})$ and
\begin{eqnarray*}
&&
\gamma_{2}(t;X_{1},\dots,X_{n})
\\
&&\quad=
\frac{1}{24 \hat{I}_n^{2}}a_{4}t^{4}+\frac{1}{72\hat{I}_n^{3}}a^{2}_{3}t^{6}+\frac{1}{2\hat{I}_n}t^{2}\frac{\pi^{\prime\prime}(\hat{\theta}_{n})}{\pi(\hat{\theta}_{n})}
\\
&&\qquad{}+
\frac{1}{6 \hat{I}_n^2}a_{3}t^{4}\frac{\pi^{\prime}(\hat{\theta}_{n})}{\pi(\hat{\theta}_{n})} - \frac{a_{4}}{8\hat{I}_n^{2}}
\\
&&\qquad{}-
\frac{15a^{2}_{3}}{72 \hat{I}_n^{3}}-\frac{1}{2\hat{I}_n}\frac{\pi^{\prime\prime}(\hat{\theta}_{n})}
{\pi(\hat{\theta}_{n})}-\frac{a_{3}}{2\hat{I}_n^2}\frac{\pi^{\prime} (\hat{\theta}_{n})}{\pi(\hat{\theta}_{n})}.
\end{eqnarray*}
\end{theorem}

The proof is given in Ghosh, Delampdy and Sa\-manta (\citeyear{gds2006}, pages
107--108). The statement involves a few minor typos which can be
corrected easily. We outline here only a few key steps needed in the
proof.

We begin with the posterior p.d.f.,
\begin{eqnarray}\label{e2p1}
&&\qquad
\pi(\theta| X_{1}, \dots, X_{n})\nonumber
\\[-8pt]\\[-8pt]
&&\qquad\quad=
\exp[\ell_{n} (\theta)] \pi(\theta) \bigl/ \int\exp[\ell_{n} (\theta)] \pi(\theta)\,d
\theta.\nonumber
\end{eqnarray}
Substituting $t= \sqrt{n} (\theta- \hat{\theta}_{n}) \hat{I}_{n}^{1/2}$,
the posterior p.d.f. of $T_{n}$ is given by
%
\begin{eqnarray}\label{e2p2}
&&\pi^{*}_{n} (t)
=
C^{-1}_{n} \exp[\ell_{n} \{\hat{\theta}_{n} +t(n \hat{I}_{n})^{-1/2}\}-\ell_{n} (\hat{\theta}_{n})]\nonumber
\\
&&\hphantom{\pi^{*}_{n} (t)=}{}\cdot
 \pi\{ \hat{\theta}_{n} + t (n
\hat{I}_{n})^{-1/2}\},\nonumber
\\[-0.5pt]
&&\hspace{56pt}\mbox{where } C_{n} = \int\exp[\ell_{n} \{\hat{\theta}_{n}
+ t (n \hat{I}_{n})^{-1/2}\}
\\[-0.5pt]
\eqntext{-\ell_{n} (\hat{\theta}_{n})]}
\\
\eqntext{\cdot\pi\{\hat{\theta}_{n} + t (n \hat{I}_{n})^{-1/2}\}\, dt.\hspace{5pt}}
\end{eqnarray}
The rest of the proof involves a~Taylor\vspace*{1pt} expansion of
$\exp[\ell_{n}\{\hat{\theta}_{n}+t(n \hat{I}_{n})^{-1/2}\}$ and
$\pi\{\hat{\theta}_{n}+t(n\hat{I}_{n})^{-1/2}\}$\break around
$\hat{\theta}_n$ up to a desired order, and collecting the coefficients
of $n^{-1/2}$, $n^{-1}$, etc. The other component is evaluation of
$C_{n}$ via momets of the N(0, 1) distribution.

\begin{remark}\label{re1}
The above result is useful in finding certain
expansions for the posterior moments as well. In particular, noting
$\th= \hat{\th}_{n} + (n \hat{I}_{n})^{-1/2} t_{n}$, it follows that
the asymptotic expansion of the posterior mean of $\th$ is given by
\begin{eqnarray}\label{e2p5}
&&
\qquad E(\th|X_{1},\dots,X_{n})\nonumber
\\[-7pt]\\[-7pt]
&&\qquad\quad=
\hat{\th}_{n}+n^{-1}\biggl\{\frac{a_{3}}{2\hat{I}_{n}^{2}}+t\frac{\pi^{\prime}(\hat{\th}_{n})}{\hat{I}_{n}\pi(\hat{\th}_{n})}\biggr\}+O_{p}(n^{-3/2}).\nonumber
\end{eqnarray}
Also, $V(\theta|X_1,\ldots,X_n)=(n\hat{I}_n)^{-1}+O_p(n^{-3/2})$.
\end{remark}

A multiparameter extension of Theorem \ref{teo1} is as follows. Suppose that
$\th= (\th_{1}, \dots, \th_{p})^{T}$ is the parameter vector and
$\hat{\th}_{n}$ is the MLE of $\th$. Let
\begin{eqnarray*}
a_{jr}
&=&
-\hat{I}_{njr} = n^{-1} \frac{\partial^{2}\ell_{n}(\th)}{\partial\th_{j}\,\partial\th_{r}}\biggl|_{\th= \hat{\th}_{n}},
\\[2pt]
a_{jrs}
&=&
n^{-1}\frac{\partial^{3}\ell_{n} (\th)}{\partial{\th}_{j}\,\partial\th_{r}\,\partial\th_{s}}\biggl|_{\th= \hat{\th}_{n}}
\end{eqnarray*}
and $\hat{I}_{n} = ((\hat{I}_{njr}))$. Then retaining only up to the
$O(n^{-1/2})$ term, the posterior of $W_{n} = \sqrt{n} (\th- \hat
{\th}_{n})$ is given by
%
\begin{eqnarray}\label{e2p6}
&&
\quad\pi^{*}_{n}(w) = (2\pi)^{-1/2} \exp[-(1/2) w^{T}\hat{I}_{n}w]\nonumber
\\[2pt]
&&\quad\hphantom{\pi^{*}_{n}(w) =}{}\cdot
\Biggl[1+ n^{-1/2}\Biggl\{\sum^{p}_{j=1}w_j\biggl(\frac{\partial\log\pi}{\partial\th_{j}}\biggr)\biggl|_{\th=\hat{\th}_{n}}\nonumber
\\[-7pt]\\[-7pt]
&&\quad\hphantom{\Biggl[1+ n^{-1/2}\Biggl\{\biggl(\frac{\partial\log\pi}{\partial\th_{j}}\biggr)\biggl|}{}+
\frac{1}{6} \sum_{j,r,s} w_{j} w_{r}w_{s}a_{jrs}\Biggr\}\nonumber
\\[2pt]
&&\hphantom{\quad\Biggl[1+ n^{-1/2}\Biggl\{\sum^{p}_{j=1}w_j\biggl(\frac{\partial\log\pi}{\partial\th_{j}}\biggr)\biggl|_{\th=\hat{\th}_{n}}}{} +
O_{p}
(n^{-1})\Biggr].\nonumber
\end{eqnarray}

Next we present the basic shrinkage argument of J.~K.~Ghosh discussed in
detail in Datta and Mukherjee (\citeyear{dm2004}). The prime objective here is
evaluation of $E[q(X,\theta)|\theta]= \lambda(\theta)$, say, where $X$
and $\theta$ can be\break real- or vector-valued. The idea is to find
first\break
$\int\lambda(\theta)\bar{\pi}_m(\theta)\,d\theta$ through a
sequence of
priors $\{\bar{\pi}_m(\theta)\}$ defined on a compact set, and then
shrinking the prior to degeneracy at some interior point, say, $\theta$
of\vadjust{\eject} the compact set. The interesting point is that one never needs
explicit specification of $\bar{\pi}_m(\theta)$ in carrying out this
evaluation. We will see several illustrations of this in this article.

First, we present the shrinkage argument in a nutshell. Consider a
proper prior $\bar{\pi} (\cdot)$ with a compact rectangle as its
support in the parameter space, and $\bar{\pi}(\cdot)$ vanishes on the
boundary of support, while remaining positive in the interior. The
support of $\bar{\pi} (\cdot)$ is the closure of the set. Consider the
posterior of $\th$ under $\bar{\pi} (\cdot)$ and, hence, obtain
$E^{\bar{\pi}} [q (X,\th) | X]$. Then find
$E[\{E^{\bar{\pi}}(q(X,\theta)|X)\}|\theta]=\lambda(\theta)$ for
$\theta$ in the interior of the support of $\bar{\pi}(\cdot)$. Finally,
integrate $\la(\cdot)$ with respect to $\bar{\pi} (\cdot)$, and then
allow $\bar{\pi} (\cdot)$ to converge to the degenerate prior at the
true value of $\th$ at an interior point of the support of~$\pi(\th)$.
This yields $E[q (X,\th)|\th]$. The calculation assumes integrability
of~$q(X,\th)$ over the joint distribution of $X$ and~$\th$. Such
integrability allows change in the order of integration.

When executed up to the desired order of approximation, under suitable
assumptions, these steps can lead to significant reduction in the
algebra underlying higher order frequentist asymptotics. The
simplification arises from two counts. First, although the Bayesian
approach to frequentist asymptotics requires Edgeworth type
assumptions, it avoids an explicit Edgeworth expansion involving
calculation of approximate cumulants. Second, as we will see, it helps
establish the results in an easily interpretable compact form. The
following two sections will de\-monstrate multiple usage of these two
basic tools.

\section{Objective Priors Via Maximization of the Distance Between the
Prior and the Posterior}\label{s3}

\subsection{Reference Priors}\label{s31}

We begin with an alternate derivation of the reference prior of
Bernardo. Following Lindley (\citeyear{l1956}), Bernardo (\citeyear{b1979}) suggested a
Kullback--Leibler (\mbox{K--L}) divergence between the prior and the posterior,
na\-mely, $E[\log\frac{\pi(\th|X)} {\pi(\th)}]$, where expectation is
taken over the joint distribution of $X$ and $\th$. The target is to
find a prior $\pi$ which maximizes the above distance. It is shown in
Berger and Bernardo (\citeyear{bb1989}) that if one does this maximization for a
fixed $n$, this may lead to a discrete prior with finitely many jumps,
a far cry from a diffuse prior. Hence, one needs an asymptotic
maximization.

First write $E[\log\frac{\pi(\th|X)}{\pi(\th)}]$ as
\begin{eqnarray}
&&
E\biggl[\log\frac{\pi(\th|X)}{\pi(\th)}\biggr]\nonumber
\\
&&\quad= \int\!\!\int
\log\frac{\pi(\th|X)}{\pi(\th)}\pi(\th|X)m^{\pi}(X)\,d\th\,
dX\nonumber
\\[-8pt]\\[-8pt]
&&\quad=
\int\!\!\int\log\frac{\pi(\th|X)}{\pi(\th)}L_{n}(\th
)\pi(\th)\,dX \,d\th\nonumber
\\
&&\quad=
\int\pi(\th) E\biggl[\log\frac{\pi(\th|X)}{\pi(\th)} \biggl| \th
\biggr]\,d \th,
\nonumber
\end{eqnarray}
where $X=(X_{1},\dots,X_{n})$, $L_{n}(\th)=\prod_{1}^{n}f(X_{i}|\th)$,
the likelihood function, and $m^{\pi}(X)$ denotes the marginal of $X$
after integrating out $\theta$. The integrations are carried out with
respect to a prior $\pi$ having a compact support, and subsequently
passing on to the limit as and when necessary.

Without any nuisance parameters, Bernardo\break (\citeyear{b1979}) showed somewhat
heuristically that Jeffreys' prior achieves the necessary maximization.
A more rigorous proof was supplied later by Clarke and Barron (\citeyear{cb1990,cb1994}). We demonstrate heuristically how the shrinkage argument can also
lead to the reference priors derived in Bernardo (\citeyear{b1979}). To this end,
we first consider the one-parameter case for a regular family of
distributions. We rewrite
\begin{eqnarray}\label{e3p8}
\quad\qquad E\biggl[\log\frac{\pi(\theta|X)}{\pi(\theta)}\biggr]
&=&
\int\pi(\theta)E[\log\pi(\theta|X)|\theta]\,d\theta\nonumber
\\[-8pt]\\[-8pt]
&&{}-
\int\pi(\theta)\log\pi(\theta)\,d\theta.\nonumber
\end{eqnarray}
Next we write
\[
E^{\bar{\pi}}[\log\pi(\theta|X)|X]=\int\log\pi
(\theta|X)
\bar{\pi}(\theta|X)\,d\theta.
\]
From the asymptotic expansion of the posterior, one gets
\begin{eqnarray*}
\log\pi(\theta|X)
&=&
(1/2)\log(n)-(1/2)\log(2\pi)
\\
&&{}-
n\frac{(\theta-\hat{\theta}_n)^2}{2}\hat{I}_n+(1/2)\log(\hat{I}_n)
\\
&&{}+
O_p(n^{-1/2}).
\end{eqnarray*}
Since $\frac{n(\theta-\hat{\theta}_n)^2}{2}\hat{I}_n$ converges
a posteriori to a $\chi_1^2$ distribution as $n\rightarrow\infty$,
irrespective of a prior $\pi$, by
the Bernstein--Von Mises and Slutsky's theorems, one gets
%
\begin{eqnarray}\label{e3p9}
&&
E^{\bar{\pi}}[\log\pi(\theta|X)]\nonumber
\\
&&\quad=
(1/2)\log(n)-(1/2)\log(2\pi e)
\\
&&\qquad{}+
(1/2)\log(\hat{I}_n)+O_p(n^{-1/2}).\nonumber
\end{eqnarray}
Since the leading term in the right-hand side of (\ref{e3p9}) does not involve
the prior $\bar{\pi}$, and $\hat{I}_n$ converges almost surely
($P_{\theta}$)
to $I(\theta)$, applying the shrinkage argument, one gets
from (\ref{e3p9})
%
\begin{eqnarray}\label{e3p10}
&&
E[\log\pi(\theta|X)|\theta]\nonumber
\\
&&\quad=
(1/2)\log(n)-(1/2)\log(2\pi e)
\\
&&\qquad{}+
\log(I^{1/2}(\theta))+O(n^{-1/2}).\nonumber
\end{eqnarray}
In view of (\ref{e3p8}), considering only the leading terms in
(\ref{e3p10}), one needs to find a prior $\pi$ which maximizes
$\int\log\{\frac{I^{1/2}(\theta)}{\pi(\theta)}\}\pi(\theta
)\,d\theta$.
The integral being nonpositive due to the property of the
Kullback--Leibler information number, its maximum value is zero, which
is attained for $\pi(\theta)=I^{1/2}(\theta)$, leading once again to
Jeffreys' prior.

The multiparameter generalization of the above result without any
nuisance parameters is based on the asymptotic expansion
\begin{eqnarray*}
&&E[\log\pi(\theta|X)|\theta]
\\
&&\quad =
(p/2)\log(n)-(p/2)\log(2\pi e)
\\
&&\qquad{}+
\int\log\{|I(\theta)|^{1/2}/{\pi(\theta)}\}\pi(\theta)\,d\theta
\\
&&\qquad{}+
O(n^{-1/2}),
\end{eqnarray*}
and maximization of the leading term yields once again Jeffreys'
general rule prior $\pi(\theta)=|I(\theta)|^{1/2}$.

In the presence of nuisance parameters, however, Jeffreys' general rule
prior is no longer the distance maximizer. We will demonstrate this in
the case when the parameter vector is split into two groups, one group
consisting of the parameters of interest, and the other involving the
nuisance parameters. In particular, Bernardo's (\citeyear{b1979}) two-group
reference prior will be included as a special case.

To this end, suppose $\theta=(\theta_1,\theta_2)$, where $\theta_1$
($p_1\times1$) is the parameter of interest and $\theta_2$ ($p_2\times
1$) is the nuisance parameter. We partition the Fisher information
matrix $I(\theta)$ as
\[
I(\theta)=\pmatrix{
I_{11}(\theta) & I_{12}(\theta)\cr
I_{21}(\theta) & I_{22}(\theta)
}.
\]
First begin with a general conditional prior\break $\pi(\theta_2|\theta_1)=
\phi(\theta)$ (say). Bernardo (\citeyear{b1979}) considered
$\phi(\theta)= |I_{22}(\theta)|^{1/2}$. The marginal prior
$\pi(\theta_1)$ for $\theta_1$ is then obtained by maximizing the
distance\break $E[\log\frac{\pi(\theta_1|X)}{\pi(\theta_1)}]$. We begin
by writing
%
\begin{equation}\label{e3p11}
\qquad\log\frac{\pi(\theta_1|X)}{\pi(\theta_1)}=
\log\frac{\pi(\theta|X)}{\pi(\theta)}-
\log\frac{\pi(\theta_2|\theta_1,X)}{\pi(\theta_2|\theta_1)}.\hspace*{-10pt}
\end{equation}
Writing $\pi(\theta)=\pi(\theta_1)\phi(\theta)$ and $|I(\theta)|=
|I_{22}(\theta)|\cdot\break|I_{11.2}(\theta)|$, where
$I_{11.2}(\theta)\!=\!
I_{11}(\theta)\!-\!I_{12}(\theta)I_{22}^{-1}(\theta)\!\cdot\! I_{21}(\theta)$,
the asymptotic expansion and the shrinkage argument together yield
\begin{eqnarray}\label{e3p12}
&&\hspace*{-8pt}
E\biggl[\log\frac{\pi(\theta|X)}{\pi(\theta)}\biggr]\hspace*{-15pt}\nonumber
\\
&&\hspace*{-8pt}\quad=
(p/2)\log(n)- (p/2)\log(2\pi e)\hspace*{-15pt}\nonumber
\\
&&\hspace*{-8pt}\qquad{}+
\int\pi(\theta_1)\hspace*{-15pt}\nonumber
\\[-8pt]\\[-8pt]
&&\hspace*{-8pt}\hphantom{\qquad{}+\int}{}\cdot
\biggl\{\int\phi(\theta)\hspace*{-15pt}\nonumber
\\
&&\hspace*{-8pt}\hphantom{\hphantom{\qquad{}+}{}\cdot
\biggl\{\int\int}{}\cdot
\log\frac{|I_{22}(\theta)|^{1/2}|I_{11.2}(\theta)|^{1/2}}{\pi(\theta_1)\phi(\theta)}\,d\theta_2\biggr\}\,d\theta_1\hspace*{-15pt}\nonumber
\\
&&\hspace*{-8pt}\qquad{}+
O(n^{-1/2})\hspace*{-15pt}\nonumber
\end{eqnarray}
and
\begin{eqnarray}\label{e3p13}
&&\quad
E\biggl[\log\frac{\pi(\theta_2|\theta_1,X)}{\pi(\theta_2|\theta_1)}\biggr]\hspace*{-10pt}\nonumber
\\
&&\quad\quad=
(p_2/2)\log(n)-(p_2/2)\log(2\pi e)\hspace*{-10pt}\nonumber
\\[-8pt]\\[-8pt]
&&\quad\qquad{}+
\int\pi(\theta_1)\biggl\{\int\phi(\theta)\log\frac{|I_{22}(\theta)|^{1/2}}{\phi
(\theta)}\,d\theta_2\biggr\}\,d\theta_1\hspace*{-10pt}\nonumber
\\
&&\quad\qquad{}+
O(n^{-1/2}).\hspace*{-10pt}\nonumber
\end{eqnarray}
From (\ref{e3p11})--(\ref{e3p13}), retaining only the leading term,
%
\begin{eqnarray}\label{e3p14}
&&\quad\quad
E\biggl[\log\frac{\pi(\theta_1|X)}{\pi(\theta_1)}\biggr]\hspace*{-5pt}\nonumber
\\
&&\quad\quad\quad\approx
(p_1/2)\log(n)-(p_1/2)\log(2\pi e)\nonumber\hspace*{-5pt}
\\[-8pt]\\[-8pt]
&&\quad\quad\qquad{}+
\int\pi(\theta_1)\hspace*{-5pt}\nonumber
\\
&&\quad\quad\hphantom{\qquad{}+\int}{}\cdot
\biggl\{\int\phi(\theta)\log\frac{|I_{11.2}(\theta)|^{1/2}}{\pi(\theta_1)}\,d\theta_2\biggr\}\,d\theta_1.\hspace*{-5pt}\nonumber
\end{eqnarray}
Writing $\log\psi(\theta_1)=\int\phi(\theta)
\log|I_{11.2}(\theta)|^{1/2}\,d\theta_2$, once again by property of
the Kullback--Leibler information number, it follows that the
maximizing prior $\pi(\theta_1)=\psi(\theta_1)$.

We have purposely not set limits for these integrals. An important
point to note [as pointed out in Berger and Bernardo (\citeyear{bb1989})] is that
evaluation of all these integrals is carried out over an increasing
sequence of compact sets $K_i$ whose union is the entire parameter
space. This is because most often we are working with improper priors,
and direct evaluation of these integrals over the entire parameter
space will simply give $+\infty$
which does not help finding any prior. As an illustration, if the parameter
space is $\mathcal{R}\times\mathcal{R}^{+}$ as is typically the case for
location--scale family of distributions, then one can take the
increasing sequence of compact sets as $[-i,i]\times[i^{-1},i]$,
$i\geq
2$. All the proofs are usually carried out by taking a sequence of
priors $\pi_i$ with compact support~$K_i$, and eventually making
$i\rightarrow\infty$. This important point should be borne in mind in
the actual derivation of reference priors. We will now illustrate this
for the location--scale family of distributions when one of the two
parameters is the parameter of interest, while the other one is the
nuisance parameter.

\begin{example}[(Location--scale models)]\label{ex1}
Suppose $X_1,\ldots,X_n$ are
i.i.d. with common p.d.f. $\sigma^{-1} f((x-\mu)/\sigma)$, where
$\mu\in(-\infty,\infty)$ and $\sigma\in(0,\infty)$. Consider the
sequence of priors $\pi_i$ with support $[-i,i]\times[i^{-1},i]$,
$i=2,3,\ldots.$ We may note that $I(\mu,\sigma)=\break\sigma^{-2}
{{c_1 \enskip c_2}\choose {c_2 \enskip c_3}}$, where the constants $c_1$, $c_2$ and $c_3$ are functions of $f$ and
do not involve either~$\mu$ or $\sigma$. So, if~$\mu$ is the
parameter
of interest, and $\sigma$ is the nuisance parameter, following
Bernardo's (\citeyear{b1979}) prescription, one begins with the sequence of
priors $\pi_{i2}(\sigma|\mu)=k_{i2}\sigma^{-1}$ where, solving $1=k_{i2}
\int_{i^{-1}}^{i}\sigma^{-1}\,d\sigma$, one gets
$k_{i2}=(2\log i)^{-1}$. Next one finds the prior
$\pi_{i1}(\mu)\!=\!k_{i1}\exp
[\int_{-i}^{i}k_{i2}\sigma^{-1}\!\log(\sigma^{-1}\!)\,d\sigma]$ which
is a constant not depending on either $\mu$ or $\sigma$. Hence, the
resulting joint prior
$\pi_i(\mu,\sigma)=\pi_{i1}(\mu)\pi_{i2}(\sigma|\mu)\propto
\sigma^{-1}$, which is the desired reference prior. Incidentally, this
is Jeffreys' independence prior rather than Jeffreys' general rule
prior, the latter being proportional to~$\sigma^{-2}$. Conversely, when
$\sigma$ is the parameter of interest and $\mu$ is the nuisance
parameter, one begins with $\pi_{i2}(\mu|\sigma)=(2i)^{-1}$ and then,
following Bernardo (\citeyear{b1979}) again, one finds $\pi_{i1}(\sigma)=\break c_{i1}
\exp[\int_{i^{-1}}^{i}(2i)^{-1}\log(1/\sigma)]\,d\mu]
\propto\sigma^{-1}$. Thus, once\break again one gets Jeffreys' independence
prior. We will see in Section \ref{s5} that Jeffreys' independence prior is a
right Haar prior, while Jeffreys' general rule prior is a left Haar
prior for the location--scale family of distributions.
\end{example}

\begin{example} [(Noncentrality parameter)] Let
$X_1,\break\ldots,X_n|\mu,\sigma$ be i.i.d. N($\mu,\sigma^2$), where $\mu$ real
and \mbox{$\sigma(\!>\!0)$} are both unknown. Suppose the parameter of interest is
$\theta=\mu/\sigma$, the noncentrality parameter. With the
reparameterization $(\theta,\sigma)$ from $(\mu,\sigma)$, the
likelihood is rewritten as
$L(\theta,\sigma)\propto\sigma^{-n}\exp[-\frac
{1}{2\sigma^2}\cdot\break
\sum_{i=1}^n(X_i-\theta\sigma)^2]$. Then the per observation Fisher~%
in\-formation matrix is given by $I(\theta,\sigma)\!=\!
{{1\phantom{00000} \enskip \theta/\sigma}\choose {\theta/\sigma \enskip (\theta^2+2)/\sigma^2}}
$. Consider once again the sequence of priors $\pi_i$ with support
$[-i,i]\times [i^{-1},i]$, $i=2,3,\ldots.$ Again, following Bernardo,
$\pi_{i2}(\sigma|\theta)=k_{i2} \sigma^{-1}$, where
$k_{i2}=(2\log i)^{-1}$. Noting that $I_{11.2}
(\theta,\sigma)=1-\theta^2/(\theta^2+2)=2/(\theta^2+2)$, one gets
$\pi_{i1}(\theta)\!=\!k_{i1}\!\exp[\int_{-i}^{i}\log(\sqrt
2/(\theta^2+2)^{1/2})\,d\sigma]\!\propto(\theta^2+2)^{-1/2}$. Hence, the
reference prior in this example is given by
$\pi_R(\theta,\sigma)\propto(\theta^2+2)^{-1/2} \sigma^{-1}$. Due to
its invariance property (Datta and Ghosh, \citeyear{dg1996}), in the original
$(\mu,\sigma$) parameterization, the two-group reference prior turns
out to be $\pi_R(\mu,\sigma)\propto\sigma^{-1}
(\mu^2+2\sigma^2)^{-1/2}$.

Things simplify considerably if $\theta_1$ and $\theta_2$ are orthgonal
in the Fisherian sense, namely, $I_{12}(\theta)=0$ (Huzurbazaar, \citeyear{huzu1950};
Cox and Reid, \citeyear{cr1987}). Then if $I_{11}(\theta)$ and $I_{22}(\theta)$
factor respectively as $h_{11}(\theta_1)h_{12}(\theta_2)$ and
$h_{21}(\theta_1)h_{22}(\theta_2)$, as a special case of a more general
result of Datta and Ghosh (\citeyear{dg1995a}), it follows that the two-group
reference prior is given by\break
$h_{11}^{1/2}(\theta_1)h_{22}^{1/2}(\theta_2)$.
\end{example}

\begin{example}
As an illustration of the above, consider the
celebrated Neyman--Scott problem (Berger and Bernardo, \citeyear{bb1992a,bb1992b}). Consider
a fixed effects one-way balanced normal ANOVA model where the number of
observations per cell is fixed, but the number of cells grows to
infinity. In symbols, let $X_{i1},\ldots,X_{ik} |\theta_i$ be mutually
independent N($\theta_i,\sigma^2)$, $k\geq2$, $i=1,\ldots,n$, all
parameters being assumed unknown. Let
\mbox{$S=\sum_{i=1}^n\sum_{j=1}^k(X_{ij}-\bar{X}_i)^2/(n(k-1))$}. Then the MLE
of $\sigma^2$ is given by $(k-1)S/k$ which converges in probability [as
$n\rightarrow\infty$ to $(k-1)\sigma^2/k$], and hence is inconsistent.
Interestingly, Jeffreys'\break prior in this case also produces an
inconsistent estimator of $\sigma^2$, but the Berger--Bernardo
reference prior does not.

To see this, we begin with Fisher Information matrix
$I(\!\theta_1,\ldots,\theta_n,\sigma\!)\!=\!
k\!\operatorname{Diag}(\!\sigma^{-2},\ldots,\sigma^{-2},(1/2)n\sigma^{-4})$. Hence,
Jeffreys' prior $\pi_J(\theta_1,\ldots,\theta_n,\sigma^2)\propto
(\sigma^2)^{-n/2-1}$ which leads to the marginal posterior
$\pi_J(\sigma^2|X)\propto(\sigma^2)^{-nk/2-1}
\exp[-n(k-1)S/(2\sigma^2)]$ of~$\sigma^2$, $X$ denoting the
entire data set. Then the posterior mean of~$\sigma^2$ is given by
$n(k-1)S/(nk-2)$, while the posterior mode is given by
$n(k-1)S/(nk+2)$. Both are inconsistent estimators of $\sigma^2$, as
these converge in probability to $(k-1)\sigma^2/k$ as
$n\rightarrow\infty$.

In contrast, by the result of Datta and Ghosh (\citeyear{dg1995a}), the two-group
reference prior $\pi_R(\theta_1,\ldots,\theta_n,\break\sigma^2)\propto
(\sigma^2)^{-1}$. This leads to the marginal posterior
$\pi_R(\sigma^2| X)\propto
(\sigma^2)^{-n(k-1)/2-1}\exp[-n(k-1)S/(2\sigma^2)]$ of
$\sigma^2$. Now the posterior mean is given by $n(k-1)S/\break(n(k-1)-2)$,
while the posterior mode is given by $n(k-1)S/(n(k-1)+2)$. Both are
consistent estimators of~$\sigma^2$.
\end{example}

\begin{example}[(Ratio of normal means)] Let $X_1$ and $X_2$ be two
independent N($\theta\mu,\mu$) random variables, where the parameter of
interest is $\theta$. This is the celebrated Fieller--Creasy problem.
The Fisher information matrix in this case is $I(\theta,\mu)=(
{{\mu^2 \enskip \mu\theta}\atop {\mu\theta \enskip 1+\theta^2}}
)$. With the transformation $\phi=\mu(1+\theta^2)^{1/2}$, one obtains
$I(\theta,\phi)= \operatorname{Diag}(\phi^2(1+\theta^2)^{-2},1)$. Again, by
Dat\-ta and Ghosh (\citeyear{dg1995a}), the two-group reference prior
$\pi_R(\theta,\phi)\propto(1+\theta^2)^{-1}$.
\end{example}

\begin{example}[(Random effects model)] This example has been visited
and revisited on several occasions. Berger and Bernardo (\citeyear{bb1992b}) first
found reference priors for variance components in this problem when
the number of observations per cell is the same. Later, Ye (\citeyear{y1994}) and
Datta and Ghosh (\citeyear{dg1995a,dg1995b}) also found reference priors for this
problem. The case involving unequal number of observations per cell was
considered by Chaloner (\citeyear{cr1987}) and Dat\-ta, Ghosh and Kim (\citeyear{dgk2002}).

For simplicity, we consider here only the case with equal number of
observations per cell. Let $Y_{ij}=m+\alpha_i+e_{ij}$, $j=1,\ldots,n,
i=1,\ldots,k$. Here $m$ is an unknown parameter, while $\alpha_i$'s and
$e_{ij}$ are mutually independent with $\alpha_i$'s i.i.d.
N($0,\sigma_{\alpha}^2$) and $e_{ij}$ i.i.d. N($0,\sigma^2$). The
parameters $m$, $\sigma_{\alpha}^2$ and $\sigma^2$ are all unknown. We
write $\bar{Y}_i=\sum_{j=1}^n Y_{ij}/n$, $i=1,\ldots,k$, and $\bar{Y}=
\sum_{i=1}^k\bar{Y}_i/k$. The minimal sufficient statistic is
($\bar{Y},T,S)$, where $T=n\sum_{i=1}^k(\bar{Y}_i-\bar{Y})^2$ and
$S=\sum_{i=1}^k\sum_{j=1}^n (Y_{ij}-\bar{Y}_i)^2$.\vspace*{2pt}

The different parameters of interest that we consider are $m$,
$\sigma_{\alpha}^2/\sigma^2$ and $\sigma^2$. The common mean $m$ is of
great relevance in meta analysis (cf. Morris and Normand, \citeyear{mn1992}). Ye
(\citeyear{y1994}) pointed out that the variance ratio $\sigma_{\alpha}^2/\sigma^2$
is of considerable interest in genetic studies. The parameter is also
of importance to animal breeders, psychologists and others. Datta and
Ghosh (\citeyear{dg1995b}) have discussed the importance of $\sigma^2$, the error
variance. In order to find reference priors for each one of these
parameters, we first make the one-to-one transformation from\vspace*{1pt}
$(m,\sigma_{\alpha}^2,\sigma^2)$ to $(m,r,u)$, where $r=\sigma^{-2}$
and $u=\sigma^2/(n\sigma_{\alpha}^2+\sigma^2)$. Thus,
$\sigma_{\alpha}^2/\sigma^2 =(1-u)/(nu)$, and the likelihood $L(m,r,u)$
can be expressed as
\begin{eqnarray*}
&&\hspace*{-2pt}L(m,r,u)
\\
&&\hspace*{-2pt}\quad=
r^{nk/2}u^{k/2}\exp[-(r/2)\{nku(\bar{Y}-m)^2+uT+S\}].
\end{eqnarray*}
Then the Fisher information matrix simplifies to $I(m,r,u)=k\operatorname{Diag}
(nru,n/(2r^2),1/(2u^2)$. From Theorem 1 of Datta and Ghosh (\citeyear{dg1995a}), it
follows now that when $m$, $r$ and $u$ are the respective
parameters of interest, while the other two are nuisance parameters, the
reference priors are given respectively by $\pi_{1R}(m,r,u)=1$,
$\pi_{2R}(m,r,u)= r^{-1}$ and $\pi_{3R}(m,r,u)=u^{-1}$.
\end{example}

\subsection{General Divergence Priors}

Next, back to the one-parameter case, we consider the more general
distance (Amari, \citeyear{a1982}; Cressie and Read, \citeyear{cr1984})
\begin{eqnarray}\label{e3p15}
&&
 \quad\quad\quad D^{\pi}=\biggl[1-E\biggl\{\frac{\pi(\th|X)}{\pi(\th)}\biggr\}^{-\bet}\biggr]\bigl/\{\bet
(1-\bet)\},\nonumber
\\[-8pt]\\[-8pt]
&&\hphantom{\hspace*{3pt}\quad \quad D^{\pi}=\biggl[1-E\biggl\{\frac{\pi(\th|X)}{\pi(\th)}\biggr\}^{-\bet}\biggr]\bigl/\{\bet
(1-\bet)\}}\bet<1,\nonumber
\end{eqnarray}
which is to be interpreted as its limit when \mbox{$\beta\ra0$}. This limit is
the K--L distance as considered in Ber\-nardo (\citeyear{b1979}). Also, $\bet= 1/2$ gives
the Bhattacharyya--Hellin\-ger (Bhattacha\-ryya, \citeyear{b1943}; Hellin\-ger, \citeyear{h1909}) distance,
and $\bet=-1$ leads to the chi-square distance (Clarke and Sun, \citeyear{cs1997,cs1999}).
In order to maximi\-ze~$D^{\pi}$ with respect to a prior $\pi$, one re-expres\-ses~(\ref
{e3p15}) as
\begin{eqnarray}\label{e3p16}
\hspace*{3pt}D^{\pi}
&=&
\biggl[1-\int\!\!\int\pi^{\bet+1} (\th) \pi^{-\bet}
(\th|X) L_{n} (\th)\,dX\,d\th\biggr]\hspace*{-25pt}\nonumber
\\
&&{}\bigl/
\{\bet(1-\bet)\} \nonumber
\\[-8pt]\\[-8pt]
&=&
\biggl[1- \int\pi^{\bet+1} (\th) E[\{\pi^{-\bet} (\th|X) \}
|\th]\,d\th\biggr]\nonumber
\\
&&{}\bigl/
\{\bet(1-\bet)\}.\nonumber
\end{eqnarray}
Hence, from (\ref{e3p16}), maximization of $D^{\pi}$ amounts to minimization
(maximization) of
%
\begin{equation}\label{e3p17}
\int\pi^{\bet+1} (\th) E[\{\pi^{-\bet} (\th|X)\}|\th]\,d\th
\end{equation}
for $0<\beta<1$ ($\beta<0$). First consider the case
$0<|\beta|\break<1$. From Theorem \ref{teo1}, the posterior of $\th$ is
\begin{eqnarray}\label{e3p18}
\qquad \pi(\th|X)
&=&
\frac{\sqrt{n}\hat{I}_{n}^{1/2}}{(2\pi)^{1/2}}\exp\biggl[-\frac{n}{2} (\th- \hat{\th}_{n})^{2} \hat{I}_{n}\biggr]\nonumber
\\[-8pt]\\[-8pt]
&&{}\cdot
[1+ O_{p}(n^{-1/2})].\nonumber
\end{eqnarray}
Thus,
%
\begin{eqnarray}\label{e3p19}
&&\hspace*{10pt}\pi^{-\beta} (\th|X)\hspace*{-16pt}\nonumber
\\
&&\hspace*{10pt}\quad=
n^{-\bet/2} (2\pi)^{\bet/2} \hat{I}_{n}^{-\bet/2}\hspace*{-16pt}
\\
&&\hspace*{10pt}\qquad{}\cdot
\exp\biggl[\frac{n\bet}{2}(\th-\hat{\th}_{n})^{2}\hat{I}_{n}\biggr]
[1+O_{p}(n^{-1/2})].\hspace*{-16pt}\nonumber
\end{eqnarray}
Following the shrinkage argument, and noting that conditional on
$\th$, $\hat{I}_{n}\!\stackrel{p}{\ra}\!I(\th)$, while $n(\th\!-\!\hat
{\th}_{n})^{2}
\hat{I}_{n}\!\stackrel{d}{\ra}\!\chi_1^2$, it follows heuristically from
(\ref{e3p19})
%
\begin{eqnarray}\label{e3p20}
&&\quad\quad
E[\pi^{-\bet}(\th|X)]\nonumber
\\
&&\quad\quad\quad=
n^{-\bet/2}(2\pi)^{\beta/2}[I(\th)]^{-\bet/2}
(1-\beta)^{-1/2}
\\
&&\quad\quad\qquad{}\cdot
[1+ O_{p} (n^{-1/2})].\nonumber
\end{eqnarray}
Hence, from (\ref{e3p20}), considering only the leading term, for
$0\!<\!\beta\!<\!1$, minimization of (\ref{e3p17}) with respect to~$\pi$
amounts to minimization of $\int[\pi(\th)/I^{1/2} (\th)]^{\bet}
\pi(\th)\,d\th$ with respect to $\pi$ subject to $\int\pi(\th)\,d\th=1$. A simple application of Holder's inequality shows that this
minimization takes place when $\pi(\th) \propto I^{1/2}(\th)$.
Similarly, for $-1<\beta<0$, $\pi(\th) \propto I^{1/2}(\th)$ provides
the desired maximization of the expected distance between the prior and
the posterior. The K--L distance, that is, when $\beta\rightarrow0,$
has already been considered earlier.

\begin{remark}
Equation (\ref{e3p20}) also holds for $\beta<-1$.
However, in this case, it is shown in Ghosh, Mergel\vadjust{\goodbreak} and Liu (\citeyear{gml2010}) that the
integral $\int\{\pi(\th)/\break I^{1/2}(\th)\}^{-\bet}\cdot\pi(\th)\,d\th
$ is
uniquely minimized with respect to $\pi(\th)\propto I^{1/2}(\th)$, and
there exists no maximi\-zer of this integral when $\int\pi(\th)\,d\th=1$. Thus, in this case, there does not exist any prior which
maximizes the posterior-prior distance.
\end{remark}

\begin{remark}
Surprisingly, Jeffreys' prior is not necessarily the
solution when $\bet=-1$ (the chi-square divergence). In this case, the
first-order asymptotics does not work since $\pi^{\bet+1}(\th)=1$ for
all $\th$. However, retaining also the $O_p(n^{-1})$ term as given in
Theorem~\ref{teo1}, Ghosh, Mergel and Liu (\citeyear{gml2010}) have found in this case the
solution\vspace*{-2pt} $\pi(\th)\propto\exp[\int^{\theta}
\frac{2g_{3}(t)-I^{\prime} (t)}{4I(t)}\,dt]$, where
$g_{3}(t)=E[-\frac{d^3 \log p (X_{1}|t)}{dt^3}|t]$. We shall refer to
this prior as $\pi_{\mathrm{GML}}(\theta)$. We will show by examples that this
prior may differ from Jeffreys'prior. But first we will establish a
hitherto unknown invariance property of this prior under one-to-one
reparameterization.
\end{remark}

\begin{theorem}\label{teo2}
Suppose that $\phi$ is a one-to-one twi\-ce differentiable
function of $\theta$. Then
$\pi_{\mathrm{GML}}(\phi)=\break C\pi_{\mathrm{GML}}(\theta)|\frac{d\theta}{d\phi}|$, where
$C(>0)$, the constant of proportionality, does not involve any
parameters.
\end{theorem}

\begin{pf}
$\!\!\!$Without loss of generality, assume that~$\phi$ is a
nondecreasing function of $\theta$. By the identity
\[
g_3(\phi)=I^{\prime}(\phi)+E\biggl[\biggl(\frac{d^2\log f}
{d\phi^2}\biggr)\biggl(\frac{d\log f}{d\phi}\biggr)\biggr],
\]
$\pi_{\mathrm{GML}}^{\prime}(\phi)/\pi_{\mathrm{GML}}(\phi)$ reduces to
\begin{eqnarray}\label{e3p22}
&&
\quad\quad\pi_{\mathrm{GML}}^{\prime}(\phi)/\pi_{\mathrm{GML}}(\phi)\hspace*{-5pt}\nonumber
\\[-8pt]\\[-8pt]
&&\quad\quad\quad=
\frac{I^{\prime}(\phi)+2E[(d^2\log f/d\phi^2)(d\log f/d\phi)]}{4I(\phi)}.\hspace*{-5pt}\nonumber
\end{eqnarray}
Next, from the relation $I(\phi)=I(\theta)(d\theta/d\phi)^2$, one
gets the
identities
\begin{eqnarray}\label{e3p23}
&&\qquad I^{\prime}(\phi)
=
I^{\prime}(\theta)\biggl(\frac{d\theta}{d\phi}\biggr)^3\nonumber
\\[-8pt]\\[-8pt]
&&\qquad\hphantom{I^{\prime}(\phi)
=}{}+
2I(\theta)(d\theta/d\phi)(d^2\theta/d\phi^2);\nonumber
\\\label{e3p24}
&&
\qquad\biggl(\frac{d^2\log f}{d\phi^2}\biggr)\biggl(\frac{d\log f}{d\phi}\biggr)\nonumber
\\
&&\qquad\quad=
\biggl\{\frac{d^2\log f}{d\theta^2}\biggl(\frac{d\theta}{d\phi}\biggr)^2+\frac
{d\log f}
{d\theta}\cdot\frac{d^2\theta}{d\phi^2}\biggr\}
\\
&&\qquad\hphantom{\quad=}{}\cdot\biggl(\frac{d\log f}{d\theta}\cdot
\frac{d\theta}{d\phi}\biggr).\nonumber
\end{eqnarray}
From (\ref{e3p22})--(\ref{e3p24}), one gets, after simplification,
%
\begin{eqnarray}\label{e3p25}
&&
\pi_{\mathrm{GML}}^{\prime}(\phi)/\pi_{\mathrm{GML}}(\phi)\nonumber
\\
&&\quad=
\frac{\pi_{\mathrm{GML}}^{\prime}(\theta)}{\pi_{\mathrm{GML}}(\theta)}\frac
{d\theta}{d\phi}+
\frac{d^2\theta/d\phi^2}{d\theta/d\phi}.
\end{eqnarray}
Now, on integration, it follows from (\ref{e3p25}) $\pi_{\mathrm{GML}}(\phi)\!=\!C
\pi_{\mathrm{GML}}(\phi)(d\theta/d\phi)$, which proves the theorem.
\end{pf}

\begin{example}
Consider the one-parameter exponential family of
distributions with $p(X|\th)=\break\exp[\th X - \psi(\th)+h(X)]$. Then
$g_{3} (\th) = I^{\prime}(\th)$ so that $\pi(\th) \propto\exp
[\frac{1}{4} \int\frac{I^{\prime} (\th)}{I(\th)}\,d\th]= I^{1/4}
(\th)\vspace{2pt}$, which is different from Jeffreys' $I^{1/2}(\theta)$ prior.
Because of the invariance result proved in Theorem \ref{teo2}, in particular,
for the $\operatorname{Binomial}(n,p)$ problem, noting that
$p=\exp(\th)/\vspace*{2pt}\break[1+\exp(\th)]$, one gets $\pi_{\mathrm{GML}}(p)\propto
p^{-3/4} (1-p)^{-3/4}$, which is a $\operatorname{Beta} (\frac{1}{4},\frac{1}{4})$ prior,
different from Jeffreys' $\operatorname{Beta} (\frac{1}{2}, \frac{1}{2})$ prior,
Laplace's $\operatorname{Beta} (1,1)$ prior or Haldane's improper $\operatorname{Beta} (0,0)$ prior.
Similarly, for the Poisson $(\lambda)$ case, one gets
$\pi_{\mathrm{GML}}(\lambda)\propto\lambda^{-1/4}$, again different from
Jeffreys' $\pi_{J}(\lambda)\propto\lambda^{-1/2}$ prior. However, for
the $\mathrm{N}(\th,1)$ distribution, since $I(\th)=1$ and
$g_{3}(\th)\!=\!
I^{\prime}(\th)\!=\!0, \pi_{\mathrm{GML}}(\th)\!=\!c(>0)$, a constant, which is the same
as Jeffreys' prior. It may be pointed~out also that for the
one-parameter exponential \mbox{family}, for the chi-square divergence,
$\pi_{\mathrm{GML}}$ differs from Hartigan's (\citeyear{h1998}) maximum likelihood prior
$\pi_{H} (\th)=I(\th)$.
\end{example}

\begin{example}
For the one-parameter location fa\-mily of distributions
with $p(X| \th) \!=\! f(X- \th)$, where~$f$ is a p.d.f., both $g_{3}(\th)$ and
$I(\th)$ are constants\vadjust{\eject} implying $I^{\prime}(\th) = 0$. Hence,
$\pi_{\mathrm{GML}}(\th)$ is of the form\break $\pi_{\mathrm{GML}}(\th)=\exp(k\th)$ for some
constant $k$. However, for the special case of a symmetric $f$, that
is, $f(X)=f(-X)$ for all $X$, $g_{3} (\th) = 0$, and then
$\pi_{\mathrm{GML}}(\th)$ reduces once again to $\pi(\th)=c$, which is the same
as Jeffreys' prior.
\end{example}

\begin{example}
For the general scale family of distributions with
$p(X|\th)= \th^{-1} f (\frac{X}{\th}), \th> 0$, where $f$ is a
p.d.f., $I
(\th) = \frac{c_{1}}{\th^{2}}$ for some constant $c_{1} (> 0)$, where
$g_{3}(\th)= \frac{c_{2}}{\th^{3}}$ for some constant $c_{2}$. Then
$\pi_{\mathrm{GML}}(\th)\propto\exp(c\log\th)=\th^{c}$ for some constant $c$.
In particular, when $p(X| \th) = \th^{-1}\exp(-\frac{X}{\th}),
\pi_{\mathrm{GML}}(\th)\propto\th^{-3/2}$, different from Jeffreys'
$\pi_J(\th)\propto\theta^{-1}$ for the general scale family of
distributions.
\end{example}

The multiparameter extension of the general divergence prior has been
explored in the Ph.D. dissertation of Liu (\citeyear{l2009}). Among other things,
he has shown that in the absence of any nuisance parameters, for
$|\beta|<1$, the divergence prior is Jeffreys' prior. However, on the
boundary, namely, $\beta=-1$, priors other than Jeffreys' prior emerge.

\section{Probability Matching Priors}\label{s4}

\subsection{Motivation and First-Order Matching}\label{s41}
As mentioned in the \hyperref[s1]{Introduction}, probability ma\-tching priors are
intended to achieve Bayes-frequen\-tist synthesis. Specifically, these
priors are required to provide asymptotically the same coverage
probability of the Bayesian credible intervals with the corresponding
frequentist counterparts. Over the years, there have been several
versions of such priors-quan\-tile matching priors, matching priors for
distribution functions, HPD matching priors and matching priors
associated with likelihood ratio statistics. Datta and Mukerjee
provided a detailed account of all these priors. In this article I will
be concerned only with quantile matching priors.

A general definition of quantile matching priors is as follows: Suppose
$X_{1}, \dots, X_{n}| \th$ i.i.d. with common p.d.f. $f (X|\th)$, where
$\theta$ is a real-valued parameter. Assume all the needed regularity
conditions for the asymptotic expansion of the posterior around~%
$\hat{\th}_{n}$, the MLE of $\th$. We continue with the notation of the
previous section. For $ 0 < \alpha< 1$, let $\th^{\pi}_{1-\alpha}
(X_{1}, \dots,\break X_{n}) \equiv\th^{\pi}_{1-\alpha}$ denote the
$(1-\alpha)$th asymptotic posterior quantile of $\th$ based on the
prior $\pi$, that is,
\begin{eqnarray}\label{e4p1}
&&
P^{\pi}[\th\leq\th^{\pi}_{1- \alpha}|X_{1},\dots,X_{n}]\nonumber
\\[-8pt]\\[-8pt]
&&\quad=
1-\alpha+O_{p}(n^{-r})\nonumber
\end{eqnarray}
for some $r\!>\! 0$. If now $P[\th\!\leq\!\th^{\pi}_{1-\alpha}|\th
]\!=\!1\!-\!\alpha
+O_{p}(n^{-r})$, then some order of probability matching is achieved.
If $r=1$, we call $\pi$ a first-order probability matching prior. If
$r= 3/2$, we call $\pi$ a second-order probability matching prior.

We first provide an intuitive argument for why Jeffreys' prior is a
first-order probability matching prior in the absence of nuisance
parameters. If $X_{1}, \dots,\break X_{n} | \th$ i.i.d. $\mathrm{N}(\th, 1)$ and
$\pi(\th) = 1$, $-\infty<\th<\infty$, then the posterior
$\pi(\th|X_{1},\dots,X_{n})$ is $\mathrm{N}(\bar{X}_{n},n^{-1})$. Now writing
$z_{1- \alpha}$ as the $100(1- \alpha)\%$ quantile of the $\mathrm{N}(0,1)$
distribution, one gets
\begin{eqnarray}\label{e4p2}
&&\qquad P\bigl[\sqrt{n}(\th-\bar{X}_{n})\leq z_{1- \alpha}|X_{1}, \dots,X_{n}\bigr)\nonumber
\\[-8pt]\\[-8pt]
&&\qquad\quad=
1-\alpha=P\bigl[\sqrt{n}(\bar{X}_{n}- \th)\geq-z_{1- \alpha}|\th\bigr],\nonumber
\end{eqnarray}
so that the one-sided credible interval $\bar{X}_{n}+z_{1-\alpha
}/\sqrt{n}$ for $\theta$ has exact frequentist coverage probability
\mbox{$1-\alpha$}.

The above exact matching does not always hold. However, if
$X_{1},\dots,X_{n}| \th$ are i.i.d., then $\hat{\th}_{n}|\th$ is
asym\-ptotically $\mathrm{N}(\theta, (nI(\th))^{-1})$. Then, by the delta me\-thod,
$g(\hat{\th}_{n})|\th\!\sim\! \mathrm{N}[g(\th),\! (g^{\prime}(\th)\!)^{2}(nI
(\th)\!)^{-1}]$. So if $g^{\prime}(\th)\!=\! I^{1/2} (\th)$ so that
$g(\th) \!=\!
\int^{\th}I^{1/2}(t)\,dt$, $\sqrt{n}[g(\hat{\th}_{n})\!-\! g(\th
)]|\th$
is asym\-ptotically $\mathrm{N}(0,1)$. Hence, from (\ref{e4p2}), with the uniform
prior $\pi( \phi)=1$ for $\phi=g(\th)$, coverage matching is
asym\-ptotically achieved for $\phi$. This leads to the prior
$\pi(\th)=\frac{d\phi}{d\th}=g^{\prime}(\th)=I^{1/2}(\th)$ for
$\th$.

Datta and Mukerjee (\citeyear{dm2004}, pages 14--21) proved the result in a formal
manner. They used the two basic tools of Section \ref{s3}. In the absence of
nuisance parameters, they showed that a first-order matching prior for
$\theta$ is a solution of the differential equation
%
\begin{equation}\label{e4p3}
\frac{d}{d\theta}(\pi(\theta)I^{-1/2}(\theta))=0,
\end{equation}
so that Jeffreys' prior is the unique first-order matching prior.
However, it does not always satisfy the second-order matching property.

\subsection{Second-Order Matching}

In order that the matching is accomplished up to $O(n^{-3/2})$ (second-order matching), one needs an asymptotic expansion of the posterior
distribution function up to the $O(n^{-1})$ term, and to set up a
second differential equation in addition to (\ref{e4p3}). This equation
is given by (cf. Mukerjee and Dey, \citeyear{md1993}; Mukerjee and Ghosh, \citeyear{mg1997})
\begin{eqnarray}\label{e4p8}
&&\qquad \frac{1}{3}\frac{d}{d\th}[\pi(\th)I^{-2}(\th)g_3(\theta)]+
\frac{d^{2}}{d\th^{2}}[\pi(\th)I^{-1}(\th)]\hspace*{-10pt}\nonumber
\\[-8pt]\\[-8pt]
&&\qquad\quad = 0,\hspace*{-10pt}\nonumber
\end{eqnarray}
where,\vspace*{1pt} as before, $g_3(\th)=-E[\frac{d^{3} \log f(X|\th)}{d\th^{3}} |
\th]$. If Jeffrey's prior satisfies (\ref{e4p8}), then it is the unique
second-order matching prior. While for the location and scale family of
distributions, this is indeed the case, this is not true in general. Of
course, in such an instance, there does not exist any second-order
matching prior.

To see this, for $\pi_{J}(\th)=I^{1/2}(\th)$, (\ref{e4p8}) reduces to
\[
\frac{1}{3} \frac{d}{d\th} [I^{-3/2}(\th)g_3(\theta)]+\frac{d^{2}}
{d\th^{2}}[I^{-1/2} (\th)]=0,
\]
which requires
$\frac{1}{3}I^{-3/2}(\th)g_3(\theta)+\frac{d}{d\th}(I^{-1/2}(\th
))$ to
be a~constant free from $\th$. After some algebra, the above expression
simplifies to
$(1/6)E[(\frac{d\log f}{d\theta})^3|\theta]/I^{3/2}(\th)$. It is
easy to check now that for the one-parameter location and scale family
of distributions, the above expression does not depend on $\theta$.
However, for the one-parameter exponential family of distributions with
canonical parameter $\theta$, the same holds if and only if
$I^{\prime}(\theta)/I^{3/2}(\theta)$ does not depend on $\theta$, or,
in other words, $I(\theta)=\exp(c\theta)$ for some constant $c$.
Another interesting example is given below.

\begin{example}
$(X_{1},X_{2})^{T}\sim \mathrm{N}_{2}[{{0}\choose{0}}
,
{{ 1 \enskip \rho }\choose { \rho \enskip 1 }}
]$. One can verify that $I(\rho)=(1+\rho^2)/(1-\rho^2)^2$ and
$L_{1,1,1}=-\frac{2\rho(3+\rho^2)}{(1-\rho^2)^{3}}$ so that
$L_{1,1,1}/I^{3/2}(\rho)$ is not a constant. Hence, $\pi_{J}$ is not a
second-order matching prior, and there does not exist any second-order
matching prior in this example.
\end{example}

\subsection{First-Order Quantile Matching Priors in the Presence
of Nuisance Parameters}

The parameter of interest is still real-valued, but there may be one or
more nuisance parameters. To fix ideas, suppose $\th=(\th_{1}, \dots,
\th_{p})$, where $\th_{1}$ is the parameter of interest, while
$\th_{2}, \dots, \th_{p}$ are the nuisance parameters. As shown by
Welch and Peers (\citeyear{wp1963}) and later more rigorously by Datta and
Ghosh\break (\citeyear{dgjk1995a}) and Datta (\citeyear{d1996}), writing $I^{-1} = ((I^{jk}))$, the
probability matching equation is given by
%
\begin{eqnarray}
\sum^{p}_{j=1}\frac{\partial}{\partial\th_{j}}\{\pi(\th
)I^{j1}(I^{11})^{-1/2}\}
= 0.
\label{e4p9}
\end{eqnarray}

\setcounter{example}{0}
\begin{example}[(Continued)]
First consider $\mu$ as the parameter of
interest, and $\sigma$ the nuisance parameter. Since each element of
the inverse of the Fisher information matrix is a constant multiple of
$\sigma^2$, any prior $\pi(\mu,\sigma)\propto g(\sigma)$, $g$
arbitrary, satisfies (\ref{e4p9}). Conversely, when $\sigma$ is the
parameter of interest, and $\mu$ is the nuisance parameter, any prior
$\pi(\mu,\sigma)\propto\sigma^{-1}g(\mu)$ satisfies (\ref{e4p9}).

A special case considered in Tibshirani (\citeyear{t1989}) is of interest. Here
$\th_{1}$ is orthogonal to $(\th_{2}, \dots, \th_{p})$ in the Fisherian
sense, that is, $I^{j1}=0$ for $j=2,3,\dots,p$.

With orthogonality, (\ref{e4p9}) simplifies to
\[
\frac{\partial}
{\partial\th_1} \{ \pi(\th) I^{-1/2}_{11} \} = 0
\]
(since $I^{11} =
I^{-1}_{11}$). This leads to $\pi(\th) =
I^{1/2}_{11}h(\th_{2},\dots,\break \th_{p})$, where $h$ is arbitrary. Often a
second-order matching prior removes the arbitrariness of $h$. We will
see an example later in this section. However, this need not always be
the case, and, indeed, as seen earlier in the one parameter case,
second-order matching priors may not always exist. We will address this
issue later in this section.

A special choice is $h\!\equiv\!1$. The resultant prior
$\pi(\theta)\!=\!I_{11}^{1/2}$ bears some intuitive appeal. Since under
ortho\-gonality, $\sqrt{n}(\hat{\th}_{1n}-\th_{1})|\th\sim
\mathrm{N}(0,I^{-1}_{11}(\th))$, one may expect $I^{1/2}_{11} (\th)$ to be a
first-order probability matching prior. This prior is only a member
within the class of priors $\pi(\th)=I^{1/2}_{11} h (\th_{2}, \dots,
\th_{p})$, as found by Tibshirani (\citeyear{t1989}), and admittedly need not be
second-order matching even when the latter exists. A recent article by
Staicu and Reid (\citeyear{sr2008}) has proved some interesting properties of the
prior $\pi(\th)= I^{1/2}_{11} (\th)$. This prior is also considered in
Ghosh and Mukerjee (\citeyear{gm1992}).

For a symmetric location--scale family of distributions, that is, when
$f(X)=f(-X)$, $c_2=0$, that is, $\mu$ and $\sigma$ are orthogonal. Now,
when $\mu$ is the parameter of interest and $\sigma$ is the nuisance
parameter, the class of first-order matching priors $\pi_1(\mu,\sigma)$
is characterized by $h_1(\sigma)$, where $h_1$ is arbitrary.
Similarly, when~$\sigma$ is the parameter of interest and $\mu$ is the
nuisance parameter, the class of first-order matching priors is
characterized by $\pi_2(\mu,\sigma)=\sigma^{-1}h_2(\mu)$, where $g_2$
is arbitrary. The intersection of the two classes leads again to the
unique prior $\pi(\mu,\sigma)=\sigma^{-1}$.
\end{example}

\begin{example}[(Continued)]
Let $X_1,\dots,X_n|\mu,\sigma$ be i.i.d.
N($\mu,\sigma^2$), and $\theta=\mu/\sigma$ is again the parameter of
interest. In order to find a parameter $\phi$ which is orthogonal to
$\theta$, we rewrite the p.d.f. in the form
\begin{eqnarray}
&&
\qquad f(X|\th, \sigma)\nonumber
\\[-8pt]\\[-8pt]
&&\qquad\quad
= (2\pi\sigma^2)^{-1/2}\exp\biggl[-\frac{1}{2\sigma^{2}}
(X-\th\sigma)^{2}\biggr].\nonumber
\end{eqnarray}
Then the Fisher information matrix
\[
I(\theta,\sigma)=\left[
\pmatrix{1 & {\theta/\sigma}\cr \theta/\sigma &
\sigma^{-2}(\theta^2+2)}
\right].
\]
 It turns out now if we reparameterize from $(\theta,\sigma)$ to
$(\theta,\phi)$, where $\phi=\sigma(\theta^2+2)^{1/2}$, then $\theta$
and $\phi$ are orthogonal with the corresponding Fisher information
matrix given by $I(\theta,\phi)=\operatorname{Diag}[2(\theta^2+2)^{-1},
\phi^{-2}(\theta^2+2)]$. Hence, the class of first-order matching
priors when $\theta$ is the parameter of interest is given by\break
$\pi(\theta,\phi)=(\theta^2+2)^{-1/2}h(\phi)$, where $h$ is arbitrary.
\end{example}

\subsection{Second-Order Quantile Matching Priors in the Presence of
Nuisance Parameters}

When $\theta_1$ is the parameter of interest, and
$(\theta_2,\ldots,\break\theta_p)$ is the vector of nuisance parameters, the
general class of second-order quantile matching priors is characterized
in (2.4.11) and (2.4.12) of Datta and Mukerjee (\citeyear{dm2004}, page 12). For
simplicity, we consider only the case when $\theta_1$ is orthogonal to
$(\theta_2,\ldots,\break\theta_p)$. In this case a first-order quantile
matching prior $\pi(\theta_1,\theta_2,\ldots,\theta_p) \propto
I_{11}^{1/2}(\theta)h(\theta_2,\ldots,\theta_p)$ is also se\-cond-order
matching if and only if $h$ satisfies (cf. Datta and Mukerjee, \citeyear{dm2004},
page 27) the differential equation
%
\begin{eqnarray}\label{e4p10}
&&
\sum_{s=2}^p\sum_{u=2}^p\frac{\partial}{\partial\theta_u}
\biggl\{I_{11}^{-1/2}I^{su}E\biggl(\frac{\partial^3\log f}
{\partial\theta_1^2\,\partial\theta_s}\biggr)h\Bigl|\theta\biggr\}\nonumber
\\
&&\quad{}+
(h/6)\frac{\partial}{\partial\theta_1}\biggl\{I_{11}^{-3/2}
E\biggl(\biggl(\frac{\partial\log f}{\partial\theta_1}\biggr)^3\Bigl|\theta\biggr)\biggr\}
\\
&&\qquad=0.\nonumber
\end{eqnarray}

We revisit Examples 1--5 and provide complete, or at least partial,
characterization of second-order quantile matching priors.

\setcounter{example}{0}
\begin{example}[(Continued)]
Let $f$ be symmetric so that $\mu$ and
$\sigma$ are orthogonal. First let $\mu$ be the parameter of interest
and $\sigma$ the nuisance parameter. Then since both the terms in
(\ref{e4p10}) are zeroes, every first-order quantile matching prior of
the form $\sigma^{-1}h(\sigma)=q(\sigma)$, say, is also second-order
matching. This means that an arbitrary prior of the form
$\pi(\mu,\sigma)$ is second-order matching as long as it is only a
function of $\sigma$. On the other hand, if $\sigma$ is the parameter
of interest and $\mu$ is the nuisance parameter, since the second term
in (\ref{e4p10}) is zero, a first-order quantile matching prior of the
form $\sigma^{-1}h(\mu)$ is also second-order matching if and only if
$h(\mu)$ is a constant. Thus, the unique second-order quantile matching
prior in this case is proportional to $\sigma^{-1}$, which is Jeffreys'
independence prior.
\end{example}

\begin{example}[(Continued)]
Recall that in this case writing
$\theta=\mu/\sigma$, and $\phi=\sigma(\theta^2+2)^{1/2}$, the Fisher
information matrix $I(\theta,\phi)\!=\!\operatorname{Diag}[2(\theta^2+2)^{-1},
\phi^{-2}(\theta^2+2)]$. Also,\vspace*{-1pt}
$E[(\frac{\partial\log f}{\partial\theta}) ^3|\theta,\phi
]\!=\!-\frac{\partial\theta(\theta^2+3)}{(\theta^2+2)^3}$ and
$E(\frac{\partial^3\log f}{\partial\theta^2
\partial\phi}|\theta,\break\phi)=(4/\phi)(\theta^2+2)^{-2}$. Hence,
(\ref{e4p10})
holds if and only if $h(\phi)=\phi^{-1}$. This leads to the unique
second-order quantile matching prior
$\pi(\theta,\phi)\propto(\theta^2+2)^{-1/2} $. Back to the
original $(\mu,\sigma)$ parameterization, this leads to the prior
$\pi(\mu,\sigma)\propto\sigma^{-1}$, Jeffreys' independence prior.
\end{example}

\begin{example}[(Continued)]
Consider once again the Neyman--Scott
example. Since the Fisher information matrix
$I(\theta_1,\ldots,\theta_n,\sigma^2)=
k\operatorname{Diag}(\sigma^{-2},\ldots,\break\sigma^{-2}, n\sigma^{-2})$,
$\sigma^2$
is orthogonal to $(\theta_1,\ldots,\theta_n)$. Now, the class of second-order matching priors is given by $\sigma^{-2}h(\mu_1,\ldots,\mu_n)$,
where $h$ is arbitrary. Simple algebra shows that in this case both the
first and second terms in (\ref{e4p10}) are zeroes so that every first-order quantile matching prior is also second-order matching.
\end{example}

\begin{example}[(Continued)]
$\!\!\!$From Tibshirani (\citeyear{t1989}), it follows that
the class of first-order quantile matching priors for $\theta$ is of
the form $(1+\theta^2)^{-1} h(\phi)$, where~$h$ is arbitrary. Once
again, since both the first and second terms in (\ref{e4p10}) are
zeroes, every first-order quantile matching prior is also second-order
matching.
\end{example}

\begin{example}[(Continued)]
Again from Tibshirani (\citeyear{t1989}), the class
of second-order matching priors when $m$, $r$ and $u$ are the parameters
of interest are given respectively by $h_1(r,u)$, $r^{-1}h_2(m,u)$ and
$u^{-1}h_3(m,r)$, where $h_1$, $h_2$ and $h_3$ are arbitrary
nonnegative functions. Also, the prior $\pi_S(r,u)\!\propto\!(ru)^{-3/2}$
is second-order matching when $m$ is the parameter of interest. On the
other hand, any first-order matching prior is also second-order
matching when either $r$ or $u$ is the parameter of interest.
\end{example}

It may be of interest to find an example where a~reference prior is not
a second-order matching prior. Consider the gamma p.d.f.
$f(x|\mu,\lambda)=(\lambda^{\lambda}/\Gamma(\lambda))\cdot
\exp[-\lambda y/\mu]y^{\lambda-1}\mu^{-\lambda}$, where the mean
$\mu$ is the parameter of interest. The Fisher \vspace {2pt}information matrix is
given by $\operatorname{Diag}(\lambda\mu^{-2},
\frac{d^2\log\Gamma(\lambda)}{d\lambda^2}- 1/\lambda)$. Then the \vspace {2pt}
two-group reference prior of Bernardo (\citeyear{b1979}) is given by
$\mu^{-1}[\frac{d^2\log\Gamma(\lambda)}{d\lambda^2}
-(1/\lambda)]^{1/2}$, while the unique second-order quantile matching
prior is given by $\lambda\mu^{-1}\cdot
[\frac{d^2\log\Gamma(\lambda)}{d\lambda^2}-(1/\lambda)]$.

In some of these examples, especially for the location and
location--scale families, one gets exact rather than asymptotic
matching. This is especially\vadjust{\eject} so when the matching prior is a
right-invariant Haar prior. We will see some examples in the next
section.

\section{Other Priors}\label{s5}

\subsection{Invariant Priors}\label{s51}

Very often objective priors are derived via some invariance criterion.
We illustrate with the location--scale family of distributions.

Let $X$ have p.d.f. $p(x|\mu,\sigma)=\sigma^{-1}f((x-\mu)/\sigma)$,
$-\infty<\mu<\infty$, $0<\sigma<\infty$, where $f$ is a p.d.f. Then, as
found in Section \ref{s4}, the Fisher information matrix $I(\mu,\sigma)$ is of
the form $I(\mu,\sigma) =\sigma^{-2}
{{c_1 \enskip c_2}\choose {c_2 \enskip c_3}} $. Hence, Jeffreys' general rule prior
$\pi_J(\mu,\sigma)\propto\sigma^{-2}$. This prior, as we will see in
this section, corresponds to a left-invariant Haar prior. In contrast,
Jeffreys' independence prior $\pi_I(\mu,\sigma)\propto\sigma^{-1}$
corresponds to a right-invariant Haar prior.

In order to demonstrate this, consider a group of linear
transformations $G=\{g_{a,b} -\infty<a<\infty,b>0\}$, where
$g_{a,b}(x)=a+bx$. The induced group of transformations on the
parameter space will be denoted by $\bar{G}$, where
$\bar{G}=\{\bar{g}_{a,b}\}$, where $\bar{g}_{a,b}(\mu,\sigma
)=(a+b\mu,
b\sigma)$. The general theory of locally compact groups states that
there exist two measures $\eta_1$ and $\eta_2$ on $\bar{G}$ such that
$\eta_1$ is left-invariant and $\eta_2$ is right-invariant. What this
means is that for all $\bar{g}\in\bar{G}$ and $A$ a subset of $G$,
$\eta_1(\bar{g}A)=\eta_1(A)$ and $\eta_2(A\bar{g})=\eta_2(A)$, where
$\bar{g}A=\{\bar{g}\bar{g_{*}}\dvtx\bar{g_{*}}\in A\}$ and
$A\bar{g}=\{\bar{g_{*}}\bar{g}\dvtx\bar{g_{*}}\in A\}$. The measures
$\eta_1$ and $\eta_2$ are referred to respectively as left- and right-invariant Haar measures. For the location--scale family of
distributions, the left- and right-invariant Haar priors turn out to be
$\pi_L(\mu,\sigma)\propto\sigma^{-2}$ and
$\pi_R(\mu,\sigma)\propto\sigma^{-1}$, respectively (cf. Berger, \citeyear{b1985},
pages 406--407; Ghosh, Delampady and Samanta, \citeyear{gds2006}, pages 136--138).

The right-Haar prior usually enjoys more optimality properties than the
left-Haar prior. Some optimality properties of left-Haar priors are
given in Datta and Ghosh (\citeyear{dgjk1995b}). In Example \ref{ex1}, for the
location--scale family of distributions, the right-Haar prior is
Bernardo's reference prior when either $\mu$ or $\sigma$ is the
parameter of interest, while the other parameter is the nuisance
parameter. Also, it is shown in Datta, Ghosh and Mukerjee (\citeyear{dgm2000}) that
for the location--scale family of distributions, the right-Haar prior
yields \textit{exact} matching of the coverage probabilities of Bayesian
credible intervals and the corresponding frequentist confidence
intervals when either $\mu$ or $\sigma$ is the parameter of interest,
while the other parameter is the nuisance parameter.

For simplicity, we demonstrate this only for the normal example. Let
$X_1,\ldots,X_n|\mu,\sigma^2$ be i.i.d. N($\mu,\break\sigma^2)$, where
$n\geq
2$. With the right-Haar prior $\pi_2(\mu,\break\sigma)\propto\sigma^{-1}$,
the marginal posterior distribution of $\mu$ is Student's $t$ with
location parameter $\bar{X}=\sum_{i=1}^n X_i/n$, scale parameter
$S/\sqrt n$, where $(n-1)S^2= \sum_{i=1}^n(X_i-\bar{X})^2$, and degrees
of freedom $n-1$. Hence, if $\mu_{1-\alpha}$ denotes the
$100(1-\alpha)$th percentile of this marginal posterior, then
\begin{eqnarray*}
1-\alpha
& = & P(\mu\leq\mu_{1-\alpha}|X_1,\ldots,X_n)
\\
& =&
P\bigl[\sqrt n(\mu-\bar{X})/S
\\
&&\hphantom{P\bigl[}{}\leq
\sqrt n(\mu_{1-\alpha}-\bar{X})/S|X_1,\ldots,X_n\bigr]
\\
& = & P\bigl[t_{n-1}\leq\sqrt n(\mu_{1-\alpha}-\bar{X})/S\bigr],
\end{eqnarray*}
so that $\sqrt n(\mu_{1-\alpha}-\bar{X})/S=t_{n-1,1-\alpha}$, the
$100(1-\alpha)$th percentile of $t_{n-1}$. Now
\begin{eqnarray*}
&&
P(\mu\leq\mu_{1-\alpha}|\mu,\sigma)
\\
&&\quad =
P\bigl[\sqrt n(\bar{X}-\mu)/S\geq-t_{n-1,1-\alpha}|\mu,\sigma\bigr]=1-\alpha
\\
&&\quad =
P(\mu\leq\mu_{1-\alpha}|X_1,\ldots,X_n).
\end{eqnarray*}
This provides the exact coverage matching probability for $\mu$.

Next, with the same set up, when $\sigma^2$ is the parameter of
interest, its marginal posterior is Inverse $\operatorname{Gamma} ((n-1)/2,
(n-1)S^2/2$). Now, if $\sigma_{1-\alpha}^2$ denotes the
$100(1-\alpha)$th percentile of this marginal posterior, then
$\sigma_{1-\alpha}^2=(n-1)S^2/\chi_{n-1;1-\alpha}^2$, where
$\chi_{n-1;1-\alpha}^2$ is the $100(1-\alpha)$th percentile of the
$\chi_{n-1}^2$ distribution. Now
\begin{eqnarray*}
&&
P(\sigma^2\leq\sigma_{1-\alpha}^2|\mu,\sigma)
\\
&&\quad=
P[(n-1)S^2/\sigma
^2\leq
\chi_{n-1;1-\alpha}^2|\mu,\sigma]=1-\alpha,
\end{eqnarray*}
showing once again the exact coverage matching.

The general definition of a right-invariant Haar density on
$\bar{\mathcal{G}}$ which we will denote by $h_r$ must satisfy
$\int_{A\bar{g}_0}h_r(x)\,dx= \int_{A}h_r(x)\,dx$, where
$A\bar{g}=\{\bar{g}_*\bar{g}\dvtx\bar{g}_*\in A\}$. Similarly, a left
invariant Haar density on $\bar{\mathcal{G}}$ which we will denote by $h_l$
must satisfy $\int_{\bar{g}A}h_l(x)\,dx\!= \int_{A}h_l(x)\,dx$, where
$\bar{g} A=\{\bar{g}\bar{g}_*\dvtx\bar{g}_*\in A\}$. An alternate
representation of the right- and left-Haar densities are given by
$P^{h_r}(A\bar{g})=P^{h_r}(A)$ and $P^{h_l}(\bar{g}A)=\break P^{h_l}(A)$,
respectively.

It is shown in Halmos (\citeyear{h1950}) and Nachbin (\citeyear{n1965}) that the right- and left-invariant Haar densities exist and are unique up to a multiplicative
constant. Berger (\citeyear{b1985}) provides calculation of $h_r$ and $h_l$ in
a~very general framework. He points out that if $\bar{\mathcal{G}}$ is
isomorphic to the parameter space $\Theta$, then one can construct
right- and left-invariant Haar priors on the parameter space $\Theta$. A
very substantial account of invariant Haar densities is available in
Datta and Ghosh (\citeyear{dgjk1995b}). Severini, Mukerjee and Ghosh (\citeyear{smg2002}) have
demonstrated the exact matching property of right invariant Haar
densities in a prediction context under fairly general conditions.

\subsection{Moment Matching Priors}

Here we discuss a new matching criterion which we will refer to as the
``moment matching criterion.'' For a regular family of distributions, the
classic article of Bernstein and Von Mises (see, e.g., Ferguson, \citeyear{f1996},
page 141; Ghosh, Delampady and Samanta, \citeyear{gds2006}, page~104) proved the
asymptotic normality of the posterior of a parameter vector centered
around the maximum likelihood estimator or the posterior mode and
variance equal to the inverse of the observed Fisher information matrix
evaluated at the maximum likelihood estimator or the posterior mode. We
utilize the same asymptotic expansion to find priors which can provide
high order matching of the moments of the posterior mean and the
maximum likelihood estimator. For simplicity of exposition, we shall
primarily confine ourselves to priors which achieve the matching of the
first moment, although it is easy to see how higher order moment
matching is equally possible.

The motivation for moment matching priors stems from several
considerations. First, these priors lead to posterior means which share
the asymptotic optimality of the MLE's up to a high order. In
particular, if one is interested in asymptotic bias or MSE reduction of
the MLE's through some adjustment, the same adjustment applies directly
to the posterior means. In this way, it is possible to achieve
Bayes-frequentist synthesis of point estimates. The second important
aspect of these priors is that they provide new viable alternatives to
Jeffreys' prior even for real-valued parameters in the absence of
nuisance parameters motivated from the proposed criterion. A third
motivation, which will be made clear later in this section, is that
with moment matching priors, it is possible to construct credible
regions for parameters of interest based only on the posterior mean and
the posterior variance, which match the maximum likelihood based
confidence intervals to a~high order of approximation. We will confine
ourselves primarily to regular families of distributions.

Let $X_1, X_2,\ldots, X_n | \theta$ be independent and identically
distributed with common density function\break $f(x|\theta)$, where $\theta
\in\Theta$, some interval in the real line. Consider a general class
of priors $\pi(\theta),\theta\in\Theta$ for $\theta$. Throughout, it
is assumed that both $f$ and $\pi$ satisfy all the needed regularity
conditions as given in Johnson (\citeyear{j1970}) and Bickel and Ghosh (\citeyear{bg1990}).

Let $\hat{\theta}_n$ denote the maximum likelihood estimator of
$\theta$. Under the prior $\pi$, we denote the posterior mean of
$\theta$ by $\hat{\theta}^B_n$. The formal asymptotic expansion given
in Section \ref{s2} now leads to\vspace*{-2pt}
$\hat{\theta}^B_n=\hat{\theta}_n+n^{-1}(\frac{a_3}{2\hat{I}_n^{2}}+
\frac{1}{\hat{I}_n}\frac{\pi^{\prime}(\hat{\theta}_n)}{\pi(\hat
{\theta}_n)}
)+O_p(n^{-3/2})$, where $a_3$ and $\hat{I}_n$ are defined in
Theorem~\ref{teo1}. The law of large numbers and consistency of the MLE now give
$n(\hat{\theta}^B_n-\hat{\theta}_n)\stackrel{P}{\rightarrow}
(\frac{-g_3(\theta)}{2I^{2}(\theta)}+\frac{1}{I(\theta)}
\frac{\pi^{\prime}(\theta)}{\pi(\theta)})$. With the choice
$\pi(\theta)=\exp[-\frac{1}{2}\int^{\theta}
\frac{g_3(t)}{I(t)}\,dt]$, one gets
$\hat{\theta}^B_n-\hat{\theta}_n= O_p(n^{-3/2})$. We will denote this
prior as $\pi_M(\theta)$.

Ghosh and Liu (\citeyear{gl2011}) have shown that if $\phi$ is a one-to-one function
of $\theta$, then the moment matching prior $\pi_M(\phi)$ for $\phi
$ is
given by $\pi_M(\phi)=\pi_M(\theta)|\frac{d \theta}{d\phi}|
^{3/2}$. We now see an application of this result.

\begin{example}[(Continued)]
Consider the regular one-parameter
exponential family of densities given by $f(x|\theta)=\exp[\theta
x-\psi(\theta)+h(x)]$. For the canonical parameter $\theta$, noting
that $I(\theta)=\psi''(\theta)$ and
$g_3(\theta)=\psi'''(\theta)=I'(\theta),$ $\pi_M(\theta
)=\exp
[\frac{1}{2}\int I'(\theta)/I(\theta)\,d\theta]=\break I^{1/2}(\theta)$,
which is Jeffreys' prior. On the other hand, for the population mean
$\phi=\psi'(\theta)$ which is a strictly increasing function of
$\theta$ [since $\psi''(\theta)=V(X|\theta) > 0]$, the moment
matching prior $\pi_M(\phi)=I(\phi)$. In particular, for the binomial
proportion $p$, one gets the Haldane prior $\pi_H(p)\propto p^{-1}
(1-p)^{-1}$, which is the same as Hartigan's (\citeyear{h1964,h1998}) maximum
likelihood prior. However, for the canonical parameter
$\theta=\operatorname{logit}(p)$, whereas we get Jeffreys' prior, Hartigan
(\citeyear{h1964,h1998}) gets the Laplace $\operatorname{uniform}(0,1)$ prior.

\begin{remark}
It is now clear that a fundamental difference between
priors obtained by matching probabilities and those obtained by
matching moments is the lack of invariance of the latter under
one-to-one reparameterization. It may be interesting to find conditions
under which a moment matching prior agrees with Jeffreys' prior
$I^{1/2}(\theta)$ or the uniform constant prior. The former holds if
and only if $g_3(\theta)=I^{\prime}(\theta)$, while the latter holds if
and only if $g_3(\theta)=0$.

The if part of the above results are immediate from the definition of
$\pi_M(\theta)$. To prove the only if parts, note that if
$\pi_M(\theta)=I^{1/2}(\theta)$, first taking logarithms, and then
differentiating with respect to~$\theta$, one gets
$\frac{I^{\prime}(\theta)}{2I(\theta)}=\frac{g_3(\theta)}
{2I(\theta)}$
so that $g_3(\theta)=I^{\prime}(\theta)$. On the other hand, if
$\pi(\theta)=c$, then taking logarithms, and then differentiating with
respect to $\theta$, one gets\break $g_3(\theta)=0$.

The above approach can be extended to the matching of higher moments as
well. Noting that $V_{\pi}(\theta|X_1,\break\ldots, X_n)
=E_{\pi}[(\theta-\hat{\theta}_n)^2|X_1,\ldots,X_n)]-(\hat{\theta}_n^B
-\hat{\theta}_n)^2$, it follows immediately that under the moment
matching prior $\pi_M$,
$V_{\pi}(\theta|X_1,\ldots,X_n)=(n\hat{I}_n)^{-1}+O_p(n^{-2})$. This
fact helps construction of credible intervals for~$\theta$, the
parameter of interest, centered at the posterior mean and scaled by the
posterior standard deviation which enjoys the same asymptotic
properties as the credible interval centered at the MLE and scaled by
the square root of the reciprocal of the observed Fisher information
number.
\end{remark}
\end{example}

\section{Summary and Conclusion}\label{s6}

As mentioned in the \hyperref[s1]{Introduction}, this article provides a selective
review of objective priors reflecting my own interest and familiarity
with the topics. I am well aware that many important contributions are
left out. For instance, I have discussed only the two-group reference
priors of Bernardo (\citeyear{b1979}). A more appealing later contribution by
Berger and Bernardo (\citeyear{bb1992b}) provided an algorithm for the construction
of multi-group reference priors when these groups are arranged in
accordance to their order of importance. In particular, the
one-at-a-time reference priors, as advocated by these authors, has
proved to be quite useful in practice. Ghosal (\citeyear{g1997,g1999}) provided the
construction of reference priors in nonregular cases, while a formal
definition of reference priors encompassing both regular and nonregular
cases has recently been proposed by Berger, Bernardo and Sun (\citeyear{bbs2009}).

Regarding probability matching priors, we have discussed only the
quantile matching criterion. There are several others, possibly equally
important probability matching criteria. Notable among these are the
highest posterior density matching criterion as well as matching via
inversion of test statistics, such as the likelihood ratio test
statistic, Rao statistic or the Wald statistic. Extensive discussion of
such matching priors is given in Datta and Mukerjee (\citeyear{dm2004}). Datta et al. (\citeyear{2000}) constructed matching priors via the
prediction criterion, and related exact results in this context are
available in Fraser and Reid (\citeyear{fr2002}). The issue of matching priors in
the context of conditional inference has been discussed quite
extensively in Reid (\citeyear{1996}).

A different class of priors called ``the maximum likelihood prior'' was
developed by Hartigan (\citeyear{h1964,h1998}). Roughly speaking, these priors are
found by maximizing the expected distance between the prior and the
posterior under a truncated Kullback--Leib\-ler distance. Like the
proposed moment matching priors, the maximum likelihood prior
densities, when they exist, result in posterior means asymptotically
negligible from the MLE's. I have alluded to some of these priors as a
comparison with other priors as given in this paper.

With the exception of the right- and left-invariant Haar priors, the
derivation of the remaining priors are based essentially on the
asymptotic expansion of the posterior density as well as the shrinkage
argument of J. K. Ghosh. This approach provides a nice unified tool for
the development of objective priors. I believe very strongly that many
new priors will be found in the future by either a direct application
or slight modification of these tools.

The results of this article show that Jeffreys' prior is a clear winner
in the absence of nuisance parameters for most situations. The only
exception is the chi-square divergence where different priors may
emerge. But that corresponds only to one special case, namely, the
boundary of the class of divergence priors, while Jeffreys' prior
continues its optimality in the interior. In the presence of nuisance
parameters, my own recommendation is to find two- or multi-group
reference priors following the algorithm of Berger and Bernardo
(\citeyear{bb1992a}), and then narrow down this class of priors by finding their
intersection with the class of probability matching priors. This
approach can even lead to a unique objective prior in some situations.
Some simple illustrations are given in this article. I also want to
point out the versatility of reference priors. For example, for
nonregular models, Jeffreys' general rule prior does not work. But as
shown in Ghosal (\citeyear{g1997}) and Berger, Bernardo and Sun (\citeyear{bbs2009}), one can
extend the definition of reference priors to cover these situations as
well.

The examples given in this paper are purposely quite simplistic to aid
understanding mainly of readers not familiar at all with the topic.
Quite rightfully, they can be criticized as somewhat stylized. Both
reference and probability matching priors, however, have been developed
for more complex problems of practical importance. Among others, I may
refer to Berger and Yang (\citeyear{by1994}), Berger, De Oliveira and Sanso (\citeyear{2001}),
Ghosh and Heo (\citeyear{gh2003}), Ghosh, Carlin and Srivastava (\citeyear{gcs1994}) and Ghosh,
Yin and Kim (\citeyear{gyk2003}). The topics of these papers include time series
models, spatial models and inverse problems, such as linear calibration
and problems in bioassay, in particular, slope ratio and parallel line
assays. One can easily extend this list. A very useful source for all
these papers is Bernardo (\citeyear{b2005}).

\section*{Acknowledgments}
This research was supported in part by NSF Grant Number SES-0631426 and
NSA Grant Number\break MSPF-076-097. The comments of the Guest Editor and a
reviewer led to substantial improvement on the manuscript.

\end{document}